\documentclass[aps,prl,twocolumn,groupedaddress,showpacs]{revtex4}
\usepackage{graphicx}

\begin{document}

\title{Spin current shot noise as a probe of interactions in 
mesoscopic systems}

\author{O. Sauret}
\author{D. Feinberg}
\email[]{feinberg@grenoble.cnrs.fr}
\affiliation{Laboratoire d'Etudes des Propri\'et\'es Electroniques
des Solides, Centre National de la Recherche Scientifique$^{¤}$, BP 166, 
38042 Grenoble Cedex
9, France}

\date{\today}

\begin{abstract}
It is shown that the spin resolved current shot noise
can probe attractive or repulsive interactions in mesoscopic systems. 
This is illustrated in two physical situations : i) a  
normal-superconducting junction where
 the spin current noise is found to be zero, and ii) a single electron 
transistor (SET), where
the spin current noise is found to be Poissonian. Repulsive interactions 
may 
also lead to weak attractive 
correlations (bunching of opposite spins) in conditions far from 
equilibrium. Spin current shot noise 
can also be used to measure the spin relaxation 
time $T_1$, and a set-up is proposed in a quantum dot geometry.
\end{abstract}

\pacs{72.70.+m, 72.25.-b, 74.45.+c, 73.23.Hk}

\maketitle

Non-equilibrium noise plays a key role in 
mesoscopic physics \cite{BB}. Low-temperature correlations of the time 
fluctuations of the electronic current 
indeed give unique information about the charge 
and the statistics of quasiparticles. For non-interacting electrons, the 
scattering approach 
is very powerful for a variety of systems 
\cite{BB,scattering1,scattering2}. The reduction of shot noise from the 
Schottky value 
originates from the Pauli exclusion principle which 
forbids two wavepackets with the same quantum numbers to be superimposed 
\cite{MartinLandauer}. 
On the other hand, Coulomb interactions also act in correlating 
wavepackets, and noise is indeed more sensitive to interactions than the 
conductance. 
Coulomb interactions may decrease or increase noise correlations, 
depending on the 
physical regimes \cite{martin_buttiker,superpoisson,loss_sukho}. Full 
counting statistics are also promising 
as a probe of interactions \cite{fullcounting}. 
Yet, in a given mesoscopic structure, the effects on the shot noise 
of Fermi statistics and of interactions are intimately mixed. 
In contrast, we propose in this Letter that spin-resolved shot noise 
can unambiguously probe the effects of electronic interactions. The basic 
idea is that the Pauli 
principle acts only on electrons with the same spin. Therefore currents 
wavepackets carried by quasiparticles with opposite spins are only correlated 
by the interactions. 

Spin-resolved shot noise has received very little attention up to now, 
contrarily to the total current 
shot noise. For instance, spin shot noise was recently considered in 
absence of charge current \cite{wang}.
On the other hand, the effect of a spin-polarized current on 
charge and spin noise was investigated, with complex behaviours due to spin accumulation
\cite{Bulka}. Noise is also an efficient probe for testing quantum 
correlations in two-electron spin-entangled states 
\cite{loss_entangle,entanglermetal,fazio,balents} or electron spin
teleportation \cite{TP}.
In contrast, we consider here simple and general mesoscopic structures in 
which the average current is {\it not} 
spin-polarized, but where the currents carried by quasiparticles with different 
spins can be separately measured. A possible set-up for this purpose will be described at the end of 
this Letter.
To clarify our statement, let us first consider a general 
 mesoscopic device made of a normal metal with non-interacting electrons, 
non magnetic terminals $i,j$,.., 
and one channel for simplicity (generalization is obvious). In absence of 
magnetic fields and 
 spin scattering of any kind, the scattering matrix is diagonal in the 
spin variable 
 and spin-independent, 
 $s_{ij}^{\sigma\sigma'}=\delta_{\sigma\sigma'}s_{ij}$. This trivially 
 leads to spin-independent averaged currents $\langle I_i^{\sigma} \rangle 
 = \langle I_i^{-\sigma} \rangle$. In a 
 similar way, spin-resolved noise, defined as 
 $S_{ij}^{\sigma\sigma'}(t-t')=
 \frac{1}{2}\langle 
 \Delta I_i^{\sigma}(t)\Delta I_j^{\sigma'}(t')+\Delta 
 I_j^{\sigma'}(t')\Delta I_i^{\sigma}(t) \rangle$,
 where $\Delta 
 I_i^{\sigma}(t)=I_i^{\sigma}(t)-\langle I_i^{\sigma} \rangle$, can be 
 evaluated. One easily 
 finds that at any frequency the noise power between terminals $i$ and 
 $j$ is diagonal in the spin variables, 
 $S_{ij}^{\sigma\sigma'}(\omega)=\delta_{\sigma\sigma'}S_{ij}(\omega)$. 
 Thus, choosing an arbitrary spin axis $\bf z$, the total (charge) current 
 noise $S_{ij}^{ch} = S_{ij}^{\uparrow \uparrow} + S_{ij}^{\downarrow 
\downarrow}
 + S_{ij}^{\uparrow \downarrow} + S_{ij}^{\downarrow \uparrow}$ and the 
 {\it spin current noise} $S_{ij}^{sp} = S_{ij}^{\uparrow \uparrow} + 
S_{ij}^{\downarrow \downarrow}
 - S_{ij}^{\uparrow \downarrow} - S_{ij}^{\downarrow \uparrow}$, defined 
 as the correlation of the spin currents 
$I_{i}^{sp}(t)=I_i^{\uparrow}(t)-I_i^{\downarrow}(t)$,  
 are strictly equal. On the contrary, in presence of Coulomb interactions, 
one expects that $S_{ij}^{\uparrow 
 \downarrow}= S_{ij}^{\downarrow \uparrow} \neq 0$, or equivalently 
$S_{ij}^{sp} \neq S_{ij}^{ch}$. This can happen 
for instance if the scattering matrix couples carriers with opposite 
spins, as Andreev scattering at a normal-
superconductor (NS) interface, or in quantum dots in presence of strong 
Coulomb repulsion. 
The sign of the correlation $S_{ij}^{\downarrow \uparrow}$ (bunching or 
antibunching) 
is of special interest.
 
Let us first consider a NS junction, where S is a singlet superconductor 
and N a normal metal. The scattering 
matrix coupling electron (e) and holes (h) 
quasiparticles in the
metal is composed of spin-conserving normal elements $s_{ee}^{\sigma 
\sigma}$, $s_{hh}^{\sigma \sigma}$, and Andreev elements
$s_{eh}^{\sigma -\sigma}$, $s_{he}^{\sigma -\sigma}$ coupling opposite 
spins. The calculation of the total zero-
frequency noise $S^{ch}=\sum_{\sigma \sigma'}S^{\sigma \sigma'}$, using 
the unitarity of the scattering matrix, reduces at
zero temperature to the
well-known result $S^{ch}=\frac{4e^3V}{\pi \hbar} 
Tr[s_{he}^{\dagger}s_{he}(1-s_{he}^{\dagger}s_{he})]$, 
where the trace is made on the channel indexes 
\cite{beenakkerNS,martinNS}.  
We remark here that it is easy to calculate the spin-resolved correlations 
$S^{\sigma \sigma}$ and $S^{\sigma -\sigma}$, 
and to check that they are exactly equal. 
The result of this observation is that for a NS junction, at $T = 0$, the 
spin current shot noise 
is strictly zero, $S^{sp}=0$. The current correlation between electrons 
with opposite spins is $S^{\uparrow \downarrow}=S^{\uparrow \uparrow}$,
 therefore {\it positive}. This "bunching" of opposite spins carriers is 
an obvious consequence of the Andreev process, since 
each spin-up quasiparticle crossing the junction is accompanied by a 
spin-down quasiparticle. This nearly instantaneous correlation 
is due to the conversion of Cooper pair wavepackets in S, into pairs of 
normal wavepackets which carry no spin, therefore the spin current
noise is zero. It has been discussed in a three-terminal geometry in Ref. 
\onlinecite{FazioSFF}. 

\begin{figure}
\includegraphics[scale=.5]{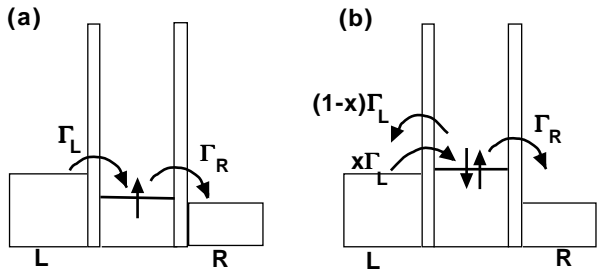}
\caption{\label{fig:SET}The SET transport sequence a) Between charge 
states $N=0$ and $1$ : rates 
$\Gamma_L$ from reservoir $L$ and $\Gamma_R$ to reservoir $R$; b) Between 
charge 
states $N=1$ and $2$ : rates 
$x\Gamma_L$ from reservoir $L$, $(1-x)\Gamma_L$ to reservoir $L$ and 
$\Gamma_R$ 
to reservoir $R$.} 
\end{figure}

Let us now turn to a very different situation, that of a quantum dot in 
the Coulomb blockade regime. Here, instead of the attractive 
correlations manifested by the NS junction, repulsive correlations are 
expected.  
 Let us consider a small island connected by tunnel 
 barriers to normal leads $L$ and $R$ with electrochemical potentials 
 $\mu_{L,R}$, such as $eV=\mu_{L}-\mu_{R}$ (Fig. \ref{fig:SET}). The 
spectrum of this quantum dot is 
 supposed to be discrete, e.g. the couplings $\Gamma_{L,R} \sim 
 2\pi|t_{L,R}|^2 N_{L,R}(0)$ to the leads verify $\Gamma_{L,R} << 
\delta\varepsilon$, 
 the level spacing. We also assume that max$(eV,k_BT) >> \hbar 
 \Gamma_{L,R}$ and that only one level of energy $E_0$ sits between 
$\mu_{R}$ 
 and $\mu_{L}$. The dot can be in three possible charge states, depending 
on 
whether the level is occupied by zero, one or two electrons (Fig. 
\ref{fig:SET}). These states will be indexed as $N=0$,
$N=1$ (with spins $\uparrow$, $\downarrow$) and $N=2$. Let us denote as 
$U(N)$ the Coulomb energy 
for the state $N$, $\Delta E_{L,R}^{+}(N)=E_0-\mu_{L,R} + U(N+1)-U(N)$ the 
energy to add an electron to state $N$ from 
leads $L,R$, and $\Delta E_{L,R}^{-}(N)=-E_0+\mu_{L,R} + U(N-1)-U(N)$ the 
energy to remove an electron from state 
$N$ towards $L,R$. Let us further assume that $\Delta E_{L}^{+}(0)$, 
$\Delta E_{R}^{-}(1)<< -k_BT$. This implies that
the transitions from $N=0$ to $1$ involve electrons coming only from $L$, 
and the transitions from $N=1$ to $0$ involve 
electrons going only into $R$. Let us allow the Coulomb energy to vary and 
consider the 
possibility of transitions from $N=1$ to $2$, only from $L$, e.g. 
$\Delta E_{R}^{-}(2)<< -k_BT$. Yet, $\Delta E_{L}^{+}(1)$ can take any 
value. This describes the following situation : if  
$\Delta E_{L}^{+}(1) >> k_BT$, the transition to state $N=2$ is forbidden 
and one has the simple SET case with 
only two charge states $N=0$, $1$, in the resonant regime at low 
temperature (Fig. \ref{fig:SET}a). 
If on the contrary $\Delta E_{L}^{+}(1) << -k_BT$, then the three charge 
states $0$, $1$, $2$ are involved in 
the charge transport (Fig. \ref{fig:SET}b).
The physical situation under consideration corresponds for instance to 
fixing the gate voltage such as $U(1)=U(0)$, and varying the 
ratio between $k_BT$ and the Coulomb excess energy $U(2)-U(1)$.   

Let us write the master equation describing this system 
\cite{mastereq1,mastereq2,mastereq3}. Assuming a constant density of 
states in the reservoirs and 
defining $x$ as the Fermi function $x = [1+exp(\beta \Delta 
E_{L}^{+}(1))]^{-1}$, 
the non-zero transition rates are $\Gamma_L^{+}(0)=\Gamma_L$, 
$\Gamma_R^{-}(1)=\Gamma_R$, $\Gamma_L^{+}(1)=x\Gamma_L$, 
$\Gamma_L^{-}(2)=(1-x)\Gamma_L$ and $\Gamma_R^{-}(2)=\Gamma_R$ (Fig. 
\ref{fig:SET}b). Then the populations $p_0$, $p_{\uparrow}$, 
$p_{\downarrow}$
 and $p_2$ verify

\begin{eqnarray}
\label{eq:master}
 \nonumber
 &\dot{p}_0 = -2\Gamma_L\, p_0 + \Gamma_R \,(p_{\uparrow} + 
p_{\downarrow})\\
 \nonumber
 &\dot{p}_{\uparrow} = -(\Gamma_R + x\Gamma_L) \,p_{\uparrow} + \Gamma_L\, 
p_0 
 + ((1-x)\Gamma_L +\Gamma_R)\, p_2\\
 &\dot{p}_{\downarrow} = -(\Gamma_R + x\Gamma_L)\, p_{\downarrow} + 
 \Gamma_L\, p_0 
 + ((1-x)\Gamma_L +\Gamma_R)\, p_2\\
\nonumber
 &\dot{p}_2 = -2((1-x)\Gamma_L +\Gamma_R)\, p_2 + 
 x\Gamma_L \,(p_{\uparrow} + p_{\downarrow})
 \end{eqnarray}

Let us first consider the limit $x=1$. Then the transition rates from $L$ 
or into $R$ do not depend on the charge state, 
which means that this limit is equivalent to that of a resonant state 
without Coulomb charging energy. 
The solution of Eqs. (\ref{eq:master}) factorizes in this case, e.g. 
 $p(n_{\uparrow},n_{\downarrow}) = p(n_{\uparrow}) p(n_{\downarrow})$, so 
that for each spin 
 component, the probabilities $p(0)$ and $p(1)$ of empty and occupied 
states 
 verify $\dot{p}(0) = -\dot{p}(1)= -\Gamma_L p(0) + \Gamma_R p(1)$. From 
this one derives the average 
 current $\langle I \rangle = 2e\frac{\Gamma_L\Gamma_R}{\Gamma_L + 
 \Gamma_R}$ and the zero-frequency shot noise $S_{ij}(\omega = 
0)=2e\langle I 
 \rangle(1-\frac{2\Gamma_L\Gamma_R}{(\Gamma_L + 
 \Gamma_R)^2})$, independently of the couple of junctions $i,j = L,R$ 
\cite{ChengTing}. Moreover, it is simple to check that spin 
 $\uparrow$ and $\downarrow$ currents are uncorrelated, thus 
$S_{ij}^{\uparrow 
 \downarrow}= S_{ij}^{\downarrow \uparrow} = 0$, or equivalently 
$S^{sp}=S^{ch}$, as can also been derived by the 
 scattering method in the quantum coherent regime. We thus have another 
example of the general behavior for uncorrelated transport.
 
 Let us now consider the SET case $x=0$, where charge transport is 
maximally correlated. The charge noise is given by the 
 expression $S_{ij}(\omega = 0)=2e\langle I 
 \rangle(1-\frac{4\Gamma_L\Gamma_R}{(2\Gamma_L + 
 \Gamma_R)^2})$ \cite{nazarovstruben}. Apart from an effective doubling of 
the rate $\Gamma_L$, this 
 result is qualitatively similar to that obtained without interactions. 
Therefore the charge noise is not the best possible 
probe of interactions. 
 We now show that, on the contrary, the behaviour of the spin noise is 
completely different. Indeed, using the method by Korotkov 
\cite{korotkov}, we find that
 
\begin{eqnarray}
\nonumber
&S_{ij}^{\sigma \sigma}= e\langle I \rangle 
(1-\frac{2\Gamma_L\Gamma_R}{(2\Gamma_L + \Gamma_R)^2})\;\;\;
S_{ij}^{\sigma -\sigma}= -e\langle I \rangle 
\frac{2\Gamma_L\Gamma_R}{(2\Gamma_L + \Gamma_R)^2}\\
&S_{ij}^{sp}=2e\langle I \rangle
\label{eq:bruitSET}
\end{eqnarray}

 \noindent
The striking result (Eq. (\ref{eq:bruitSET})) for $S^{sp}$ resembles a 
 Poisson result and corresponds to maximal fluctuations. 
 The correlations between currents of opposite spins are 
 negative, like a partition noise. Yet spin-up and 
 spin-down channels are separated energetically rather than spatially, 
 and wavepackets with up or down spins exclude 
 each other because of interactions rather than statistics. 

The above result, obtained at zero temperature with perfect spin coherence 
inside the island, 
can be interpreted in the following way : electrons come from reservoir 
$L$ with random spins. 
Even though the average spin current is zero, each junction is 
sequentially crossed -- due to Coulomb repulsion -- 
by elementary wavepackets with well-defined but uncorrelated 
spins. This implies very short time correlations (on the scale of
tunneling through one of the barriers) therefore the spin current exhibits 
Poisson statistics. On the contrary, {\it  charge} current
wavepackets are correlated on times $\sim \hbar/\Gamma_i$, leading to the 
reduction as compared to the Poisson value. 
Notice that Eqs. (\ref{eq:bruitSET}) is a consequence of the restriction 
to two charge states : as can be easily checked, 
the analysis of the SET involving only $N=1$ and $2$ states (instead of 
$0$, $1$) yields exactly the same result. 
 
 The general solution of Eqs. (\ref{eq:master}) offers an interpolation 
between the uncorrelated and the maximally correlated regimes. 
We find that the average current is given by $\langle I 
\rangle=e\frac{2\Gamma_L \Gamma_R}{\Gamma_R + (2-x) \Gamma_L}$. The spin 
current noise 
components $S_{ij}^{\sigma \sigma'}$ (i,j=L,R) can also be calculated and 
do not depend 
on the couple $(i,j)$ of junctions chosen. The
expression for the spin noise is

\begin{equation}
S_{ij}^{sp}=2e\langle I \rangle \,(1-\frac{2x\Gamma_L \Gamma_R}{(\Gamma_R 
+ \Gamma_L)(\Gamma_R+x\Gamma_L)})
\end{equation}

\begin{figure}
\includegraphics[scale=.5]{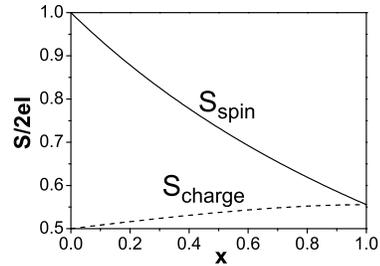}
\caption{\label{fig:bruit2}Spin shot noise and charge shot noise in the 
SET, as a function of $x$ 
(see text) : $x=0$ denotes the maximal correlation, $x=1$ the uncorrelated 
case.
$\Gamma_R=2\Gamma_L$ : antibunching of opposite spins.} 
\end{figure}

\begin{figure}
\includegraphics[scale=.5]{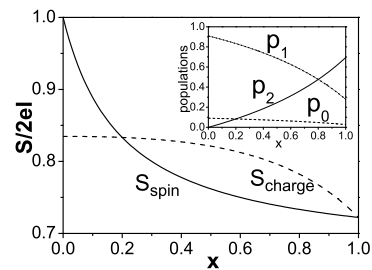}
\caption{\label{fig:bruit0.2}Same as Fig. 2, $\Gamma_R=0.2\Gamma_L$ :
bunching of opposite spins for $x>x_c$. The inset shows the probabilities 
of states $N=0,1,2$ and the population inversion at large $x$.}
\end{figure}

The expression for the total (charge) noise $S^{ch}$ is too lengthy to be 
written here. Figs.
\ref{fig:bruit2}, \ref{fig:bruit0.2} 
shows the variation with $x$ of the
charge and spin current noise. The spin noise is maximum for $x=0$, 
decreases monotonously and merges the charge noise at $x=1$. 
The role of the asymmetry of the junctions is very striking. First, if 
$\Gamma_R > \Gamma_L$, we find that $S^{sp}$ is always larger 
than $S^{ch}$ (Fig. \ref{fig:bruit2}), like in the ideal SET ($x=0$). 
On the other hand, if $\Gamma_R < \Gamma_L$, $S^{sp}$ happens to be
smaller than $S^{ch}$ for $x > x_c \sim \Gamma_R/\Gamma_L$ (Fig. 
\ref{fig:bruit0.2}). This implies that $S^{\uparrow \downarrow} > 0$, 
contrarily to the 
naive expectation for repulsive interactions. This unexpected behavior can 
be explained as follows : if $\Gamma_R < \Gamma_L$, 
the low charge states are unfavored and the high ones favored, despite of 
Coulomb repulsion :
 for $x > x_c$ , $p(2)$ becomes larger than $p(0)$. 
When the SET occasionnally reaches the state $N=0$, a first transition 
leads to state $1$, but then the most probable
transition is to state $2$ since $\Gamma_L^+(1) = x\Gamma_L > 
\Gamma_R^-(1)=\Gamma_R$ : two electrons enter the dot 
successively, with opposite spins, leading to a certain degree of 
bunching. Here the anomaly is due to a kind of 
"population inversion" (the most energetical state is favoured),  
manifesting a strong departure 
from equilibrium (Fig. \ref{fig:bruit0.2}). Yet, the effect is rather 
weak, less than $10\%$, 
contrarily to the NS junctions where attractive correlations are $100\%$.

Let us now consider how the above results are modified by spin relaxation, 
due for instance to spin-orbit scattering. 
Let us simply focus on the SET with two charge states $0$, $1$ and 
introduce a spin-flip rate $T_1^{-1}= \gamma_{sf}$.
The master equations will be written in this case 
 
 \begin{eqnarray}
 \nonumber
 &\dot{p}_0 = -2\Gamma_L\, p_0 + \Gamma_R \,(p_{\uparrow} + 
p_{\downarrow})\\
 &\dot{p}_{\uparrow} = -\Gamma_R\,p_{\uparrow} + \Gamma_L\, p_0 
 + \frac{1}{2}\gamma_{sf}\,(p_{\downarrow}-p_{\uparrow})\\
\nonumber
 &\dot{p}_{\downarrow} = -\Gamma_R\,p_{\downarrow} + \Gamma_L\, p_0 
 + \frac{1}{2}\gamma_{sf}\,(p_{\uparrow}-p_{\downarrow})
 \end{eqnarray}
 
The introduction of spin relaxation obviously changes neither the average 
curent nor the charge shot noise. On the contrary, the 
spin shot noise is altered. For instance, opposite spin noise correlations 
between junctions $L$ and $R$ become $S_{LR}^{\sigma 
 -\sigma} = -e\langle I \rangle (\frac{2\Gamma_L\Gamma_R}{(2\Gamma_L + 
 \Gamma_R)^2}-\frac{\gamma_{sf}}{2(\Gamma_R + \gamma_{sf})})$, and can 
even become positive. This results in a spin noise 

\begin{equation}
S_{LL}^{sp}=2e \langle I \rangle \; ; \;\;\; S_{LR}^{sp}=2e \langle I 
\rangle \,\frac{\Gamma_R}{\Gamma_R + \gamma_{sf}}
\end{equation}

The reduction of the spin noise $S_{LR}^{sp}$ from the "Poisson" value is 
a fingerprint of spin relaxation. Remarkably enough, 
the spin noise on
junction $L$ is not affected, since the spins of entering wavepackets are 
uncorrelated whatever 
 happens in the island. While transient current measurements have allowed 
to measure $T_1$ in presence of Zeeman 
splitting \cite{T1}, our result suggests an alternative method which does 
not require a magnetic field. Notice that noise was recently proposed 
to test spin flip in absence of Coulomb repulsion \cite{spinflip}.

\begin{figure}
\includegraphics[scale=.5]{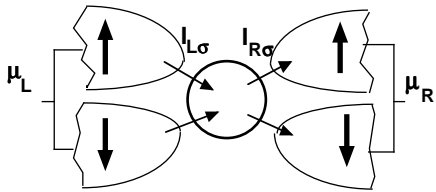}
\caption{\label{fig:4term}Schematic set-up for spin current measurement, 
using four spin-polarized terminals (see text).} 
\end{figure}

Let us now propose a set-up for the measurement of spin current 
correlations in a single-electron transistor. 
One may consider a four-terminal configuration \cite{spinflip}, where the 
two left terminals $L1$, $L2$ are ferromagnetic metals with 
opposite
spin polarizations, having the same chemical potential $\mu_L$ (Fig. 
\ref{fig:4term}). Similarly, terminals $R1$ and $R2$ have 
opposite polarizations, respectively parallel 
to those of $L1$, $L2$, and the same chemical potential $\mu_R$. If the 
junction parameters are the same for $L1$, $L2$ on one hand, and for
$R1$ and $R2$ on the other hand, then the net current flowing through the 
SET is not spin 
polarized. Yet, it is possible to
measure separately the spin current components in each of the four 
terminals, e. g. measure the noise correlations
$S_{L1L1}$, $S_{L1L2}$, $S_{L1R1}$, $S_{L1R2}$, etc... If each terminal 
generates a fully spin-polarized current,
the analysis of this set-up can be mapped onto the above model, and the 
previous results hold. In the more realistic case where polarization 
is not perfect, the
above measurement would yield a mixing of the spin noise with the charge 
noise. If those are sufficiently different 
(strong repulsive correlations), they could still be distinguished, 
which allows to probe the Coulomb correlations
 by the method of spin current noise. 

In summary, we have proposed to probe the attractive or repulsive 
correlations induced by interactions 
by measuring the noise correlations of the
spin components of the current. This requires not to break the spin 
symmetry in the device, e.g. the total current is not spin-polarized. 
We have illustrated this trend on two simple and 
classical mesoscopic devices. First, a NS
junction shows opposite spin bunching due to attractive correlations. 
Second, a SET in the sequential regime
shows in general repulsive correlations (antibunching), but those can be 
weakly attractive far from equilibrium. 
Extensions to other regimes or multiple dot systems is quite promising. 

\begin{acknowledgments}
The authors are grateful to Th. Martin for fruitful discussions concerning 
the "partition noise" analogy. 
LEPES is under convention with Universit\'e Joseph Fourier. 
\end{acknowledgments}

\end{document}